\documentclass[a4paper,11pt]{article}
\usepackage{pos}
\usepackage{subfigure}
\title{Proton and neutron electromagnetic form factors using $N_f$=2+1+1 twisted-mass fermions with physical values of the quark masses}
\ShortTitle{Proton and neutron EM form factors}

\author[a,b]{Constantia Alexandrou}
\author[b]{Simone Bacchio}
\author[b]{Giannis Koutsou}
\author*[b]{Bhavna Prasad}
\author[a]{Gregoris Spanoudes}

\affiliation[a]{Department of Physics, University of Cyprus}
\affiliation[b]{Computation-based Science and Technology Research Center, The Cyprus Institute}

\abstract{We compute the electromagnetic form factors of the proton
  and neutron using lattice QCD with $N_f = 2 + 1 + 1$ twisted mass
  clover-improved fermions and quark masses tuned to their physical
  values. Three ensembles with lattice spacings of $a$=0.080 fm, 0.068
  fm, and 0.057 fm, and approximately the same physical volume allow
  us to obtain the continuum limit directly at the physical pion
  mass. Several values of the source-sink time separation ranging from
  0.5 fm to 1.5 fm are used, enabling a thorough analysis of excited
  state effects via multi-state fits. The disconnected contributions
  are analyzed using high statistics for the two-point functions
  combined with low-mode deflation and hierarchical probing for the
  fermion loop estimation. We study the momentum dependence of the
  form factors using the z-expansion and dipole Ansaetze, thereby
  enabling the extraction of the electric and magnetic radii, as well
  as the magnetic moments in the continuum limit, for which we provide
  preliminary results.}

\FullConference{The 41st International Symposium on Lattice Field Theory (LATTICE2024)\\
 28 July - 3 August 2024\\
Liverpool, UK\\}

\begin{document}
\maketitle

\section{Introduction}

The proton and neutron electromagnetic form factors offer insights
into the rich internal electromagnetic structure of these
nucleons. Over the years, several experimental probes have
investigated these form factors, leading to a very precise
determination of the charges, moments, and radii of these
nucleons~\cite{A1:2013fsc,Punjabi:2015bba,Pohl:2010zza,Golak:2000nt,Xiong:2019umf}. In
these proceedings, we provide a calculation of the electromagnetic
form factors of the nucleon using lattice QCD on three ensembles of
clover-improved twisted mass fermions with two degenerate light,
strange, and charm quarks ($N_f=2+1+1$) with masses tuned to their
physical values (physical point). The lattice spacings span a=0.080
fm, 0.068 fm, and 0.057 fm, allowing a continuum limit directly at the
physical pion mass, while source-sink time separation ranging from 0.5
fm to 1.5 fm are used for analysis of excited states. Including
disconnected contributions, we obtain the proton and neutron electric
and magnetic form factors in the isospin limit and study their
momentum dependence to extract the electric and magnetic radii, as
well as the magnetic moments in the continuum limit.

\section{Nucleon Electromagnetic form factors}
In the flavor isospin limit, the electromagnetic form factors are
given in terms of the matrix element of the electromagnetic current
between nucleon states,
\[
\langle N(p', s')|\mathcal{O}^{V}_\mu|N(p,s)\rangle =
\sqrt{\frac{m^2_N}{E_N(\vec{p}')E_N(\vec{p})}}\bar{u}_N(p',s')\Lambda_\mu(q^2)u_N(p,s)
\]
with $N(p,s)$ a nucleon state of momentum $p$ and spin $s$,
$E_N(\vec{p}) = p_0$ its energy and $m_N$ its mass, $u_N$ a nucleon
spinor, and $q=p'-p$ the momentum transfer from initial ($p$) to final
($p'$) momentum. The matrix element is expressed in terms of the Dirac
($F_1$) and Pauli ($F_2$) form factors,
\begin{align}
  \Lambda_\mu(q^2) = \gamma_\mu
  F_1(q^2)+\frac{i\sigma_{\mu\nu}q^\nu}{2m_N}F_2(q^2),\quad
\end{align}
or alternatively in terms of the nucleon electric ($G_E$) and magnetic
($G_M$) Sachs form factors via $G_E(q^2) =
F_1(q^2)+\frac{q^2}{(2m_N)^2}F_2(q^2)$ and $G_M(q^2) =
F_1(q^2)+F_2(q^2)$. At zero momentum transfer ($q^2=0$), the electric
form factor yields the nucleon charge and the magnetic its
magnetic moment,
\begin{align}
  G_E^p(0) = 1,\,\,\,G_E^n(0)= 0,\,\,\, G_M^p(0)=\mu_p,\,\,\,\mathrm{and}\,\,\, G_M^n(0) = \mu_n,
\end{align}
where the superscripts $p$ and $n$ are used to denote the proton and
neutron form factors respectively. The electric and magnetic
root-mean-squared (r.m.s) radii are defined as the slope of the
corresponding Sachs form factor as $q^2\rightarrow 0$, namely
\begin{equation}
  \langle r_X^2 \rangle^{\mathsf{q}} = \frac{-6}{G_X^\mathsf{q}(0)}
  \frac{\partial G_X^\mathsf{q}(q^2)}{\partial q^2} \Big
  \vert_{q^2=0},\,
  \label{eq:radius}
\end{equation}
with $X=E,M$ and $\mathsf{q}=p,n$.

\section{Lattice setup}
On the lattice, we compute the nucleon three-point correlation
function,
\begin{equation}
  C_\mu(\Gamma_\nu,\vec{q},\vec{p}\,';t_s,t_{\rm ins},t_0) {=} 
  \sum_{\vec{x}_{\rm ins},\vec{x}_s}  e^{i (\vec{x}_{\rm ins} {-} \vec{x}_0)  \cdot \vec{q}}  e^{-i(\vec{x}_s {-} \vec{x}_0)\cdot \vec{p}\,'}\mathrm{Tr} [ \Gamma_\nu \langle \chi_N(x_s)  j_\mu(x_{\rm ins}) \bar{\chi}_N(x_0) \rangle],
  \label{eq:thrp}
\end{equation}
where $x_0$, $x_{\rm ins}$, and $x_s$ are referred to as the
\textit{source}, \textit{insertion}, and \textit{sink} respectively,
and $\chi_N$ is the standard nucleon interpolating
field~\cite{Alexandrou:2018sjm}. The local vector current $j_\mu$ is
given by, $j_\mu = \sum_{\mathsf{q}=u,d} e_\mathsf{q} j^\mathsf{q}_\mu
= \sum_{\mathsf{q}=u,d} e_\mathsf{q} \bar{\mathsf{q}} \gamma_\mu
\mathsf{q}$ where the sum over $\mathsf{q}$ runs over the up-
($\mathsf{q}=u$) and down- ($\mathsf{q}=d$) quark flavors and
$e_\mathsf{q}$ is the electric charge of the quark with flavor
$\mathsf{q}$. We will refer to the isovector and isoscalar flavor
combinations of the form factors, for which we use $j_\mu^{u-d} =
j_\mu^u - j_\mu^d$ and $j_\mu^{u+d} = j_\mu^u + j_\mu^d$
respectively. The twisted mass formulation we employ allows the
definition of a lattice conserved vector current which we use for the
case of the connected three-point correlation functions. $\Gamma_\nu$
is a projector acting on dirac indices, with $\Gamma_0 {=}
\frac{1}{2}(1{+}\gamma_0)$ yielding the unpolarized and
$\Gamma_k{=}\Gamma_0 i \gamma_5 \gamma_k$ the polarized matrix
elements. Without loss of generality we will take $t_s$ and $t_{\rm
  ins}$ relative to the source time $t_0$ in what follows. The
three-point function yields,
\begin{equation}
      C_{\mu}(\Gamma_\nu,\vec{q},\vec{p}';t_s,t_{\rm ins}) = 
    \sum_{n,m}  \mathcal{A}^{n,m}_{\mu}(\Gamma_\nu,\vec{q},\vec{p}') 
    e^{-E_n(\vec{p}')(t_s-t_{\rm ins})-E_m(\vec{q})t_{\rm ins}}, \label{eq:thrp_spec}
\end{equation}
where the desired ground state matrix element is
$\mathcal{A}^{0,0}_{\mu}(\Gamma_\nu,\vec{q},\vec{p}')$ multiplied by
unknown overlaps of the nucleon state with $\chi_N$. To cancel these
overlaps, we use the two-point nucleon correlation function,
\begin{align}
  C(\vec{p},t_s) = \sum_{\vec{x}_s} & e^{{-}i \vec{x}_s \cdot
    \vec{p}}\mathrm{Tr} \left[ \Gamma_0 {\langle}\chi_N(x_s)
    \bar{\chi}_N(0) {\rangle} \right] = \sum_{n}c_n(\vec{p})
  e^{-E_n(\vec{p}) t_s},\label{eq:twop_spec}
\end{align}
and form the ratio,
\begin{equation}
\label{eq:gsmatrix}
  \Pi^\mu(\Gamma_\nu;\vec{p}', \vec{q}) =
  \frac{\mathcal{A}^{0,0}_{\mu}(\Gamma_\nu,\vec{q},\vec{p}')}{\sqrt{c_0(\vec{p})
      c_0(\vec{p}')}}.
\end{equation}

\begin{table}[h]
\centering\begin{tabular}{cccccc}
  \hline\hline
  Ensemble  & $(\frac{L}{a})^3\times(\frac{T}{a})$ & $a$ [fm] & $m_\pi$ [MeV]  & $m_\pi L$ & $n_{\rm conf}$\\
  \hline 
  \texttt{cB211.072.64} & $64^3 \times 128$ &  0.07957(13) & 140.2(2) & 3.62 & 749 \\
  \texttt{cC211.060.80} & $80^3 \times 160$ &  0.06821(13) & 136.7(2) & 3.78 & 401 \\
  \texttt{cD211.054.96} & $96^3 \times 192$ &  0.05692(12) & 140.8(2) & 3.90 & 496 \\
  \hline
\end{tabular}
\caption{Parameters of the three $N_f = 2+1+1$ ensembles used. We
  provide the name of the ensemble, the lattice volume, $\beta=6/g^2$
  with $g$ the bare coupling constant, the lattice spacing, the pion
  mass, the value of $m_{\pi}L$, and the number of configurations. The
  lattice spacing values and pion masses are as obtained in
  Ref.~\cite{ExtendedTwistedMass:2022jpw}.}
  \label{tab:latticesetup}
\end{table}

We use ensembles simulated with $N_f = 2+1+1$ twisted mass,
clover-improved fermions with quark masses tuned to approximately
their physical values. A summary of the parameters for the ensembles
is provided in Table~\ref{tab:latticesetup}. The two- and three-point
functions are computed using multiple source positions per gauge
configuration. For two-point functions, we use 477, 650, and 480
source positions for the ensembles with decreasing $a$
respectively. For the connected three-point functions, we employ seven
to ten different sink-source time separations ranging from
approximately 0.5~fm to 1.5~fm with the number of source positions per
configuration increasing with separation to maintain approximately
constant statistical errors. We also compute the disconnected
contributions to the isoscalar contribution employing the
\emph{one-end trick}~\cite{McNeile:2006bz}, full dilution in color and
spin, and hierarchical probing~\cite{Stathopoulos:2013aci} to distance
eight in the 4-dimensional volume for the calculation of the fermion
loop. We also use eigenvector deflation for the two ensembles at
coarser lattice spacings. The disconnected contributions are computed using
the local vector current and therefore need to be renormalized. The
renormalization is carried out non-perturbatively in the RI'-MOM
scheme~\cite{Martinelli:1994ty} employing momentum sources, following
the procedures described in
Refs.~\cite{Alexandrou:2010me,Alexandrou:2015sea}. We refer to
Ref.~\cite{Alexandrou:2024ozj} for details on the statistics of each
sink-source separation and on the precise approach for computing the
disconnected contributions.

\section{Extraction of form factors}
The bare form factors at each value of the momentum transfer squared
($Q^2$) are obtained by appropriate combinations of $\Gamma_\nu$ and
$\mu$ depending on the momenta $\vec{p}'$ and $\vec{q}$ in
$\Pi^\mu(\Gamma_\nu;\vec{p}', \vec{q})$ of Eq.~(\ref{eq:gsmatrix}) in
order to isolate $G_E$ and $G_M$. For the connected contributions we
employ the standard \textit{fixed sink}
approach~\cite{Alexandrou:2018sjm} and therefore fix $\vec{p}'=0$. For
this case, the expressions yielding $G_E$ and $G_M$ can be
disentangled. For the disconnected contributions we combine
$\vec{p}'=\frac{2\pi}{L}\vec{k}$ for $\vec{k}^2=0$, 1, and 2. In this
case, the expressions yielding $G_E$ and $G_M$ cannot be disentangled
(see Appendix of Ref.~\cite{Alexandrou:2018sjm}) and we therefore use
a Singular Value Decomposition to solve the overconstrained set of
equations that emerge, as in the case of the Generalized Form Factors
in Ref.~\cite{Alexandrou:2019ali}.

\begin{table}[h]
    \centering\begin{tabular}{ccccccccc}
    \hline\hline
       Ensemble  & $t_s^{\rm low,3pt}$ & $t_s^{\rm low,2pt}$ &  $t_{\rm ins}^{\rm max}$ \\
       \hline 
        \texttt{cB211.072.64} & 8, 10, 12, 14 & 1, 2, 3  & 2, 3, 4 \\ 
        \texttt{cC211.060.80} & 8, 10, 12, 14 & 1, 2, 3, 4  & 2, 3, 4 \\
        \texttt{cD211.054.96} & 8, 10, 12, 14 & 1, 2, 3, 4, 5 & 2, 3, 4 \\
        \hline
    \end{tabular}
    \caption{Values of the variations used in the fit ranges for each
  ensemble. For each $t_{\rm ins}^{\rm max}$, the $t_{\rm ins}^{\rm min}$ takes values $t_{\rm ins}^{\rm max}$, $t_{\rm ins}^{\rm max}+1$ or $t_{\rm ins}^{\rm max}+2$. }
  \label{tab:fitmodels}
\end{table}
To obtain the ground-state contribution to
$\Pi^\mu(\Gamma_\nu;\vec{p}', \vec{q})$, we perform combined fits to
the two- and three-point functions. We include two excited states
(three-state fits) when fitting the two-point functions and the first
excited state (two-state fits) when fitting the three-point
function. In our fitting procedure, we first fit the two-point
functions at $\vec{p}^2=0$ and $\vec{p}^2=(\frac{2\pi}{L})^2$ to
extract the model-averaged ground-state energy, $E_0(0)$, which is
used as prior to all subsequent fits. For the ground-state energy at
finite $\vec{p}$ we use the dispersion relation throughout, namely
$E_0(\vec{p}) = \sqrt{E_0(0)^2 + \vec{p}^2}$. We proceed to fit each
value of $Q^2$, allowing for different excited state energies between
two- and three-point functions and between the connected isovector and
isoscalar cases. In these fits, we vary the smallest separation in the
two- and three-point function fits ($t_s^{\rm low,2pt}$ and $t_s^{\rm
  low,3pt}$ respectively) and the values of the insertion time
included according to $t_{\rm ins}\in[t_{\rm ins}^{\rm min},
  t_s-t_{\rm ins}^{\rm max}]$. The combinations used for each ensemble
are provided in Table~\ref{tab:fitmodels}.

\begin{figure}
  \includegraphics[width=0.495\textwidth]{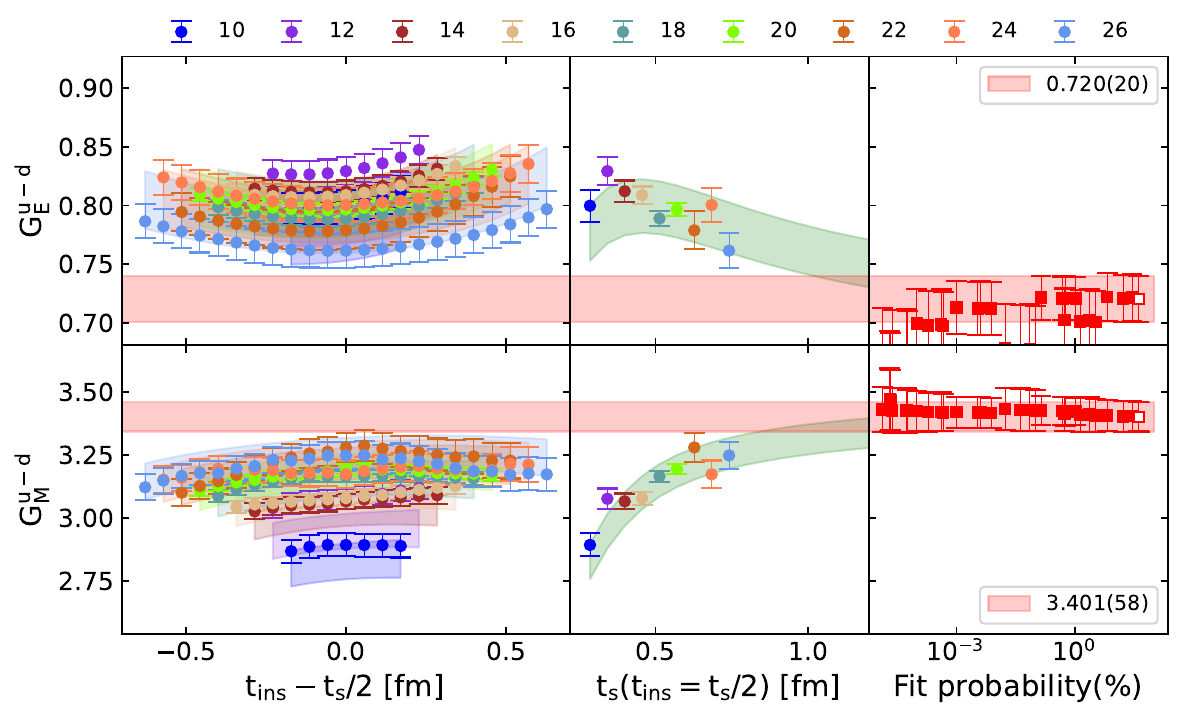}
  \includegraphics[width=0.495\textwidth]{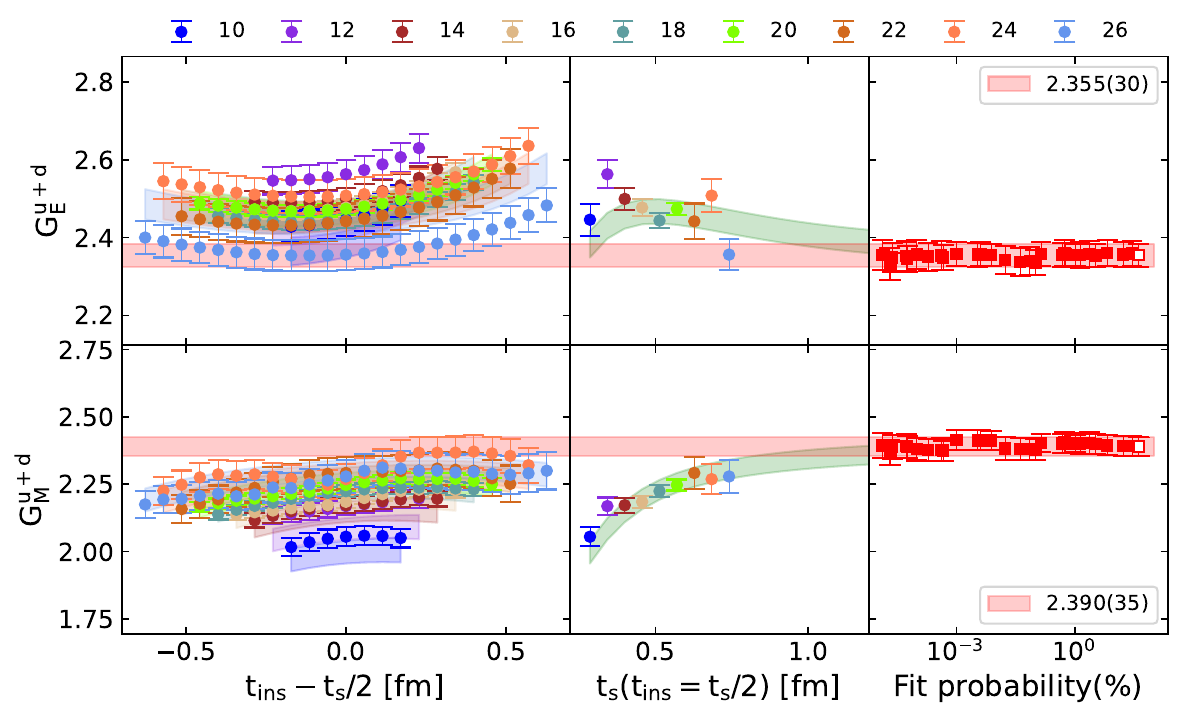}
  \caption{Extraction of the isovector (left) and isoscalar (right)
    electric (top) and magnetic (bottom) form factors for the second non-zero $Q^2$ value for the cD211.054.96 ensemble. The left
    column of each plot shows the ratio of three- to two-point
    functions described in the text for the source-sink separations
    indicated in the header of the figure. The center column shows the
    ratio for $t_{\rm ins} = t_s/2$, and the right column gives the
    result of each fit versus its fit probability. The bands show the
    most probable fit, which is also shown with the open symbol in the
    right column.}
  \label{fig:cC80_ratios}
\end{figure}

An example of this analysis is shown in Fig.~\ref{fig:cC80_ratios},
where for visualization purposes we plot the ratio of three- to
two-point functions of Ref.~\cite{Alexandrou:2018sjm}. The results for
each choice of fit ranges are model-averaged according to the Akaike
Information Criterion (AIC)~\cite{Jay:2020jkz,Neil:2022joj} following the
approach also employed in Ref.~\cite{Alexandrou:2023qbg} for the axial
form factors computed on the same ensembles.

\section{Results for form factors}
\begin{figure}
    \includegraphics[width=0.98\textwidth]{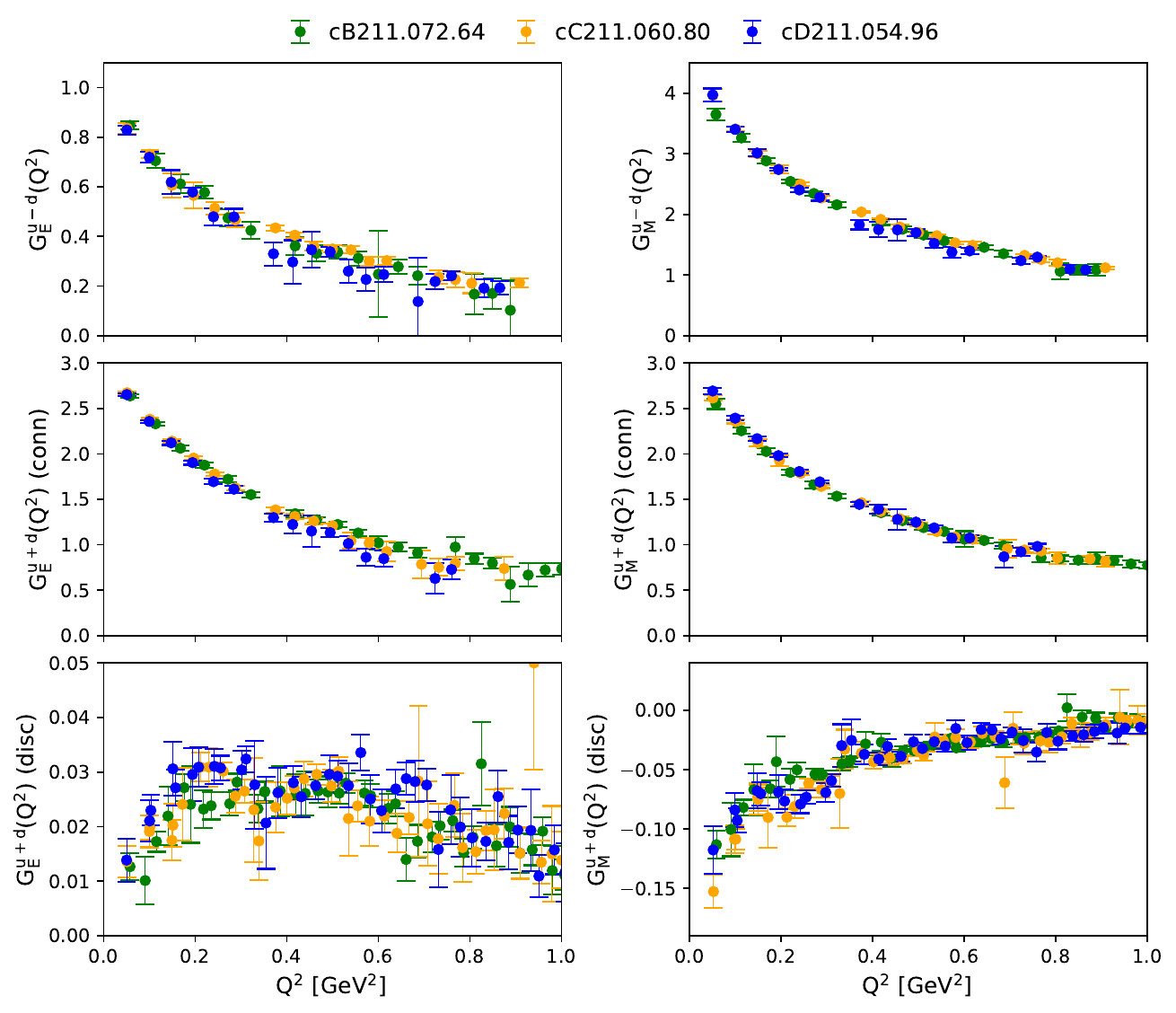}
    \caption{$G_E$ (left) and $G_M$ (right), connected isovector (top), connected isoscalar (center) and
      disconnected isoscalar (bottom) form factors as a function of $Q^2$ for the three ensembles analyzed
      here.}
    \label{fig:GEMvs}
\end{figure}

The connected and disconnected contribution to the isoscalar form
factors and the isovector form factors are shown in
Fig.~\ref{fig:GEMvs} as a function of $Q^2$ for the three ensembles
analyzed here. For each value of $Q^2$, the connected contributions
are obtained via the analysis procedure described in the previous
section. For the disconnected contributions, we do not observe
significant excited state contamination within the statistical
accuracy achieved and we therefore use results from plateau fits.
The proton and neutron form factors are obtained from the isoscalar
and isovector form factors,
\begin{equation}
  G_X^p(q^2) = \frac{1}{2}G_X^{u-d}(q^2) + \frac{1}{6}G_X^{u+d}(q^2)\,\,\,\mathrm{and}\,\,\,G_X^n(q^2) = -\frac{1}{2}G_X^{u-d}(q^2) + \frac{1}{6}G_X^{u+d}(q^2),\label{eq:vstopn}
\end{equation}
where $X=E,\,M$. We model the $Q^2$ dependence and take the continuum
limit using the proton and neutron form factors directly and fit to
both dipole and z-expansion forms. The dipole is given by
\begin{equation}
    G(Q^2) = \frac{G(0)}{1+\frac{Q^2}{M^2}},
    \label{eq:dipole}
\end{equation}
with $G(0)$ and $M^2$ the fitting parameters. The radius is obtained
via $\langle r^2\rangle = \frac{12}{M^2}$. The z-expansion is given
by
\begin{equation}
  G(Q^2) = \sum_{k=0}^{k_{max}} a_k z^k(Q^2),
  \label{eq:z-exp}
\end{equation}
with $z = \frac{\sqrt{t_{\rm cut} + Q^2} - \sqrt{t_{\rm
      cut}}}{\sqrt{t_{\rm cut} + Q^2} + \sqrt{t_{\rm cut}}}$. The
radius is $\langle r^2\rangle = - \frac{3 a_1}{2 a_0 t_{cut}}$ and we
take $t_{\rm cut} {=} (2m_{\pi})^2$. For the proton electric form
factor we fix $G(0)=1$ for the dipole and $a_0=1$ for the
z-expansion. For neutron electric form factor, we use the Galster-like
parameterization~\cite{Galster:1971kv} instead of the dipole. The
continuum limit is taken in two ways, namely i) via a
\textit{``two-step approach''}, where each ensemble's $Q^2$-dependence
is fitted separately and the radius and magnetic moment are then
extrapolated to the continuum in a second step or ii) via a
\textit{``one-step approach''}, where the $a^2$ dependence is included
in the fit of either the dipole or z-expansion and all three ensembles
are fitted together using a similar approach to that in
Ref.~\cite{Alexandrou:2023qbg}. We will quote results using both
approaches for the dipole case while for the z-expansion we use the
one-step approach. When fitting the z-expansion, we demand that the
form factor approaches zero as $Q^2\rightarrow\infty$ which fixes one
parameter, and take the order of the z-expansion such that the fit has
three free parameters, i.e. $k_{\rm max}=4$ for the case of the electric
form factors and $k_{\rm max}=3$ for the magnetic.

\begin{figure}
    \includegraphics[width=0.98\textwidth]{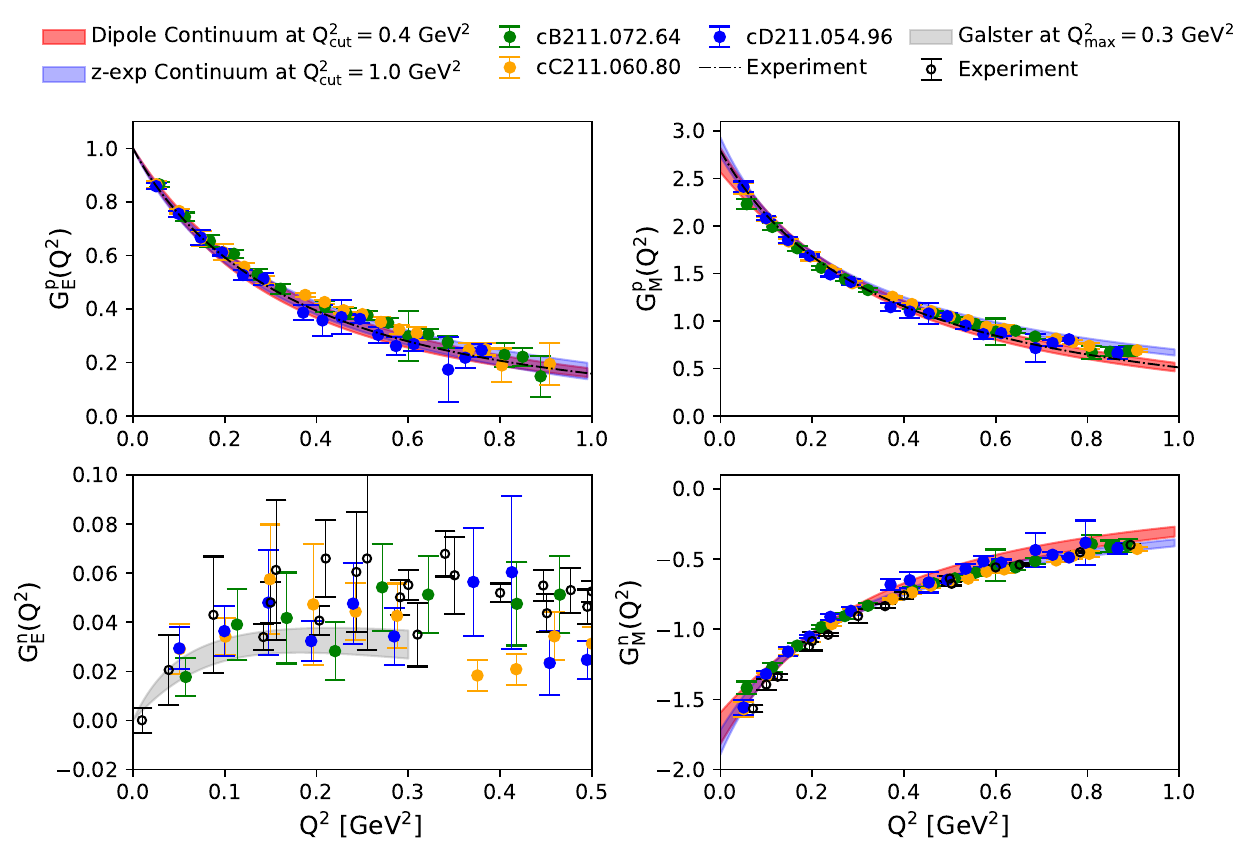}
    \caption{$G_E$ (left) and $G_M$ (right), proton (top) and neutron
      (bottom) form factors as a function of $Q^2$ for the three
      ensembles analyzed. The light red and blue bands indicate the
      continuum limit band using the dipole and z-expansion
      respectively. The black dashed lines for the proton case and the black circles for the neutron case correspond to the
      experimental results.}
    \label{fig:GEMpn}
\end{figure}

Our results for the proton and neutron electromagnetic form factors
are shown in Fig.~\ref{fig:GEMpn}, where we also show
representative continuum extrapolations using either the
z-expansion or dipole forms in the one-step approach. For the case of
the neutron electric form factor ($G^n_E(Q^2)$) the data are consistent with the experimental values within errors. However, we do not include $a^2$ dependence because of large statistical errors, and restrict the maximum value of $Q^2$ used in the fit ($Q^2_{\rm cut}$) to $Q^2_{\rm cut}$ = 0.3~GeV$^2$ as shown in Fig.~\ref{fig:GEMpn}. For the other three form factors we vary the $Q^2_{\rm cut}$ in the z-expansion and use $Q^2_{\rm cut}$ = 0.4~GeV$^2$ for the dipole. In Fig.~\ref{fig:GEMpn}, we also show representatively
$Q^2_{\rm cut}$=0.4~GeV$^2$ for dipole and $Q^2_{\rm cut}$=1~GeV$^2$
for z-expansion with $k_{\rm max}$ as explained in the previous section. The dashed black curves for the proton form
factors are from z-expansion fits to experimental
data~\cite{Ye:2017gyb}.
\begin{figure}
  \centering
    \includegraphics[width=0.8\textwidth]{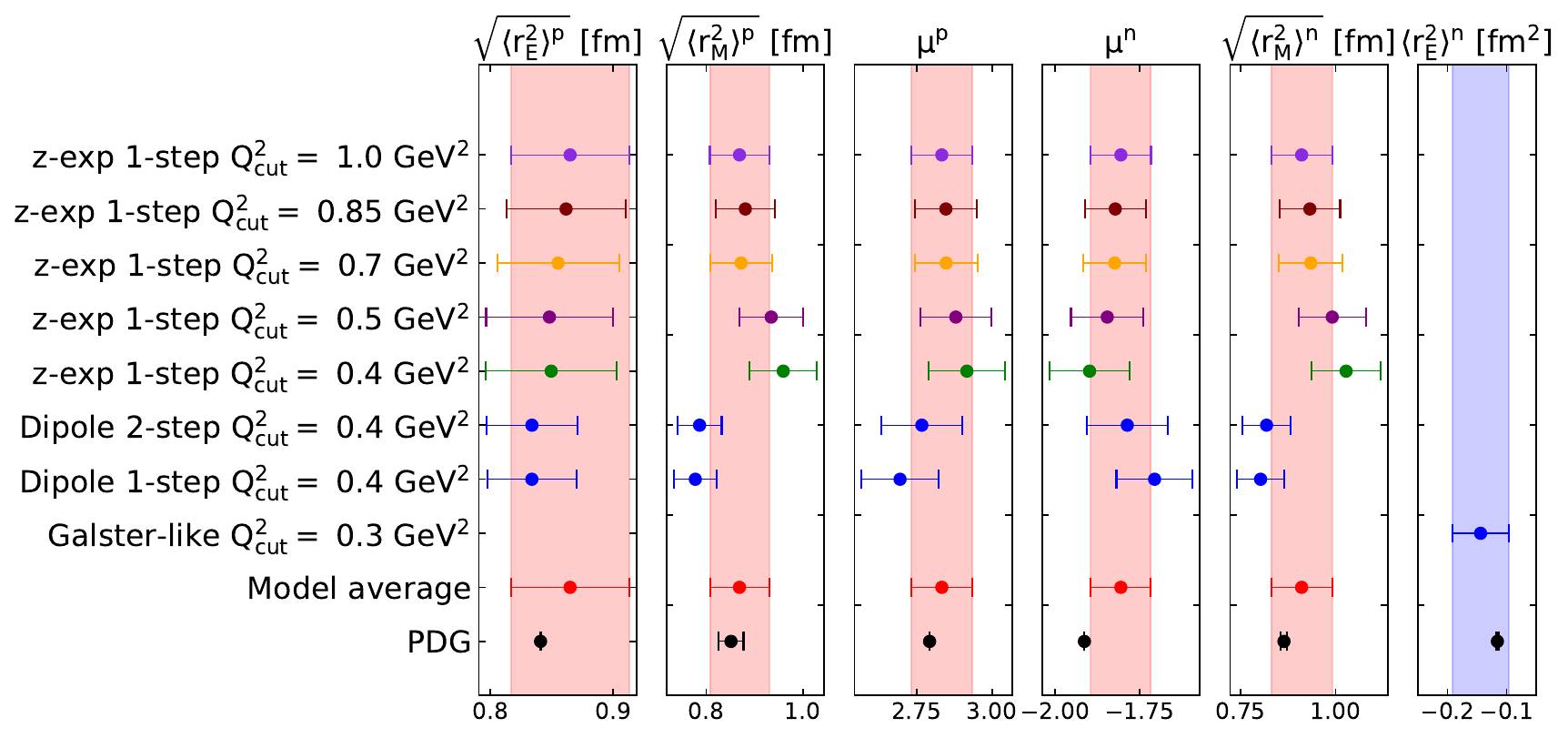}
    \caption{Electric and magnetic radii and magnetic moments of the proton and neutron for all $Q^2_{\rm cut}$ and,
      in the dipole case, using both one- or two-step approaches. The
      red point and band denoted ``Model average'' is obtained by weighting according to the AIC as explained in the
      text.}
    \label{fig:Results}
\end{figure}
Our results for the radii and magnetic moments are shown in
Fig.~\ref{fig:Results} for all $Q^2_{\rm cut}$ used and for both one-
or two-step approaches for the case of the dipole form. We overall
observe consistent results when varying the fit ansatz and the
$Q^2_{\rm cut}$ used. The model average result, also shown, is
consistent with the PDG values~\cite{ParticleDataGroup:2024cfk} for
these quantities.

\section{Conclusions}
We have carried out an analysis of the electromagnetic form factors of
the nucleon using three ensembles of $N_f=2+1+1$ twisted mass fermions
at three lattice spacings and with physical pion mass. Our excited
state analysis employs multi-state fits allowing for a different first
excited state in the two- and three-point functions and combine
multiple fit ranges via a model-average. Our results for the form factors obtained on each ensemble and the experimental results are overall compatible with each other, indicating very small cutoff effects and overall good agreement. We carry out a preliminary
continuum extrapolation in $a^2$ within a combined fit of the
$Q^2$-dependence using both dipole and z-expansion ansaetze. Our
preliminary results for the radii and magnetic moments are consistent with the PDG
values for these quantities within our combined statistical and systematic errors shown in Fig.~\ref{fig:Results}. The analysis of the excited state contamination, $Q^2$-dependence and continuum limit is continuing in order to obtain a more robust model average. We note that a fourth
ensemble with lattice spacing $a=0.049$~fm and approximately same
physical volume as the three ensembles used here is available and its analysis is ongoing, with first
results for the charges presented in Ref.~\cite{charges:2024pos}.

\section*{Acknowledgments}
C.A., G.K., and G.S.  acknowledge partial support by the projects
3D-nucleon, NiceQuarks, and ``Lattice Studies of Strongly Coupled
Gauge Theories: Renormalization and Phase Transition''
(EXCELLENCE/0421/0043, EXCELLENCE/0421/0195, and EXCELLENCE/0421/0025)
co-financed by the European Regional Development Fund and the Republic
of Cyprus through the Research and Innovation Foundation as well as
AQTIVATE that received funding from the European Union’s research and
innovation program under the Marie Sklodowska-Curie Doctoral Networks
action, Grant Agreement No 101072344. C.A acknowledges support by the
University of Cyprus projects ``Nucleon-GPDs'' and
``PDFs-LQCD''. S.B. is supported by Inno4scale, which received funding
from the European High-Performance Computing Joint Undertaking (JU) GA
No.~101118139. B.P. is supported by ENGAGE which received funding from
the EU's Horizon 2020 Research and Innovation Programme under the
Marie Skłodowska-Curie GA No. 101034267. This work was supported by
grants from the Swiss National Supercomputing Centre (CSCS) under
projects with ids s702 and s1174. The authors gratefully acknowledge
the Gauss Centre for Supercomputing e.V. (www.gauss-centre.eu) for
funding this project by providing computing time through the John von
Neumann Institute for Computing (NIC) on the GCS Supercomputer
JUWELS-Booster at J\"ulich Supercomputing Centre (JSC). The authors
also acknowledge the Texas Advanced Computing Center (TACC) at
University of Texas at Austin for providing HPC resources.

\bibliographystyle{JHEP} \bibliography{refs}

\providecommand{\href}[2]{#2}\begingroup\raggedright\begin{thebibliography}{10}

\bibitem{A1:2013fsc}
{\scshape A1} collaboration, J.~C. Bernauer et~al., \emph{{Electric and magnetic form factors of the proton}}, \href{https://doi.org/10.1103/PhysRevC.90.015206}{\emph{Phys. Rev. C} {\bfseries 90} (2014) 015206} [\href{https://arxiv.org/abs/1307.6227}{{\ttfamily 1307.6227}}].

\bibitem{Punjabi:2015bba}
V.~Punjabi, C.~F. Perdrisat, M.~K. Jones, E.~J. Brash and C.~E. Carlson, \emph{{The Structure of the Nucleon: Elastic Electromagnetic Form Factors}}, \href{https://doi.org/10.1140/epja/i2015-15079-x}{\emph{Eur. Phys. J. A} {\bfseries 51} (2015) 79} [\href{https://arxiv.org/abs/1503.01452}{{\ttfamily 1503.01452}}].

\bibitem{Pohl:2010zza}
R.~Pohl et~al., \emph{{The size of the proton}}, \href{https://doi.org/10.1038/nature09250}{\emph{Nature} {\bfseries 466} (2010) 213}.

\bibitem{Golak:2000nt}
J.~Golak, G.~Ziemer, H.~Kamada, H.~Witala and W.~Gloeckle, \emph{{Extraction of electromagnetic neutron form-factors through inclusive and exclusive polarized electron scattering on polarized He-3 target}}, \href{https://doi.org/10.1103/PhysRevC.63.034006}{\emph{Phys. Rev. C} {\bfseries 63} (2001) 034006} [\href{https://arxiv.org/abs/nucl-th/0008008}{{\ttfamily nucl-th/0008008}}].

\bibitem{Xiong:2019umf}
W.~Xiong et~al., \emph{{A small proton charge radius from an electron\textendash{}proton scattering experiment}}, \href{https://doi.org/10.1038/s41586-019-1721-2}{\emph{Nature} {\bfseries 575} (2019) 147}.

\bibitem{Alexandrou:2018sjm}
C.~Alexandrou, S.~Bacchio, M.~Constantinou, J.~Finkenrath, K.~Hadjiyiannakou et~al., \emph{{Proton and neutron electromagnetic form factors from lattice QCD}}, \href{https://doi.org/10.1103/PhysRevD.100.014509}{\emph{Phys. Rev. D} {\bfseries 100} (2019) 014509} [\href{https://arxiv.org/abs/1812.10311}{{\ttfamily 1812.10311}}].

\bibitem{ExtendedTwistedMass:2022jpw}
{\scshape Extended Twisted Mass} collaboration, C.~Alexandrou et~al., \emph{{Lattice calculation of the short and intermediate time-distance hadronic vacuum polarization contributions to the muon magnetic moment using twisted-mass fermions}}, \href{https://doi.org/10.1103/PhysRevD.107.074506}{\emph{Phys. Rev. D} {\bfseries 107} (2023) 074506} [\href{https://arxiv.org/abs/2206.15084}{{\ttfamily 2206.15084}}].

\bibitem{McNeile:2006bz}
{\scshape UKQCD} collaboration, C.~McNeile and C.~Michael, \emph{{Decay width of light quark hybrid meson from the lattice}}, \href{https://doi.org/10.1103/PhysRevD.73.074506}{\emph{Phys. Rev. D} {\bfseries 73} (2006) 074506} [\href{https://arxiv.org/abs/hep-lat/0603007}{{\ttfamily hep-lat/0603007}}].

\bibitem{Stathopoulos:2013aci}
A.~Stathopoulos, J.~Laeuchli and K.~Orginos, \emph{{Hierarchical Probing for Estimating the Trace of the Matrix Inverse on Toroidal Lattices}}, \href{https://doi.org/10.1137/120881452}{\emph{SIAM J. Sci. Comput.} {\bfseries 35} (2013) S299} [\href{https://arxiv.org/abs/1302.4018}{{\ttfamily 1302.4018}}].

\bibitem{Martinelli:1994ty}
G.~Martinelli, C.~Pittori, C.~T. Sachrajda, M.~Testa and A.~Vladikas, \emph{{A General method for nonperturbative renormalization of lattice operators}}, \href{https://doi.org/10.1016/0550-3213(95)00126-D}{\emph{Nucl. Phys. B} {\bfseries 445} (1995) 81} [\href{https://arxiv.org/abs/hep-lat/9411010}{{\ttfamily hep-lat/9411010}}].

\bibitem{Alexandrou:2010me}
C.~Alexandrou, M.~Constantinou, T.~Korzec, H.~Panagopoulos and F.~Stylianou, \emph{{Renormalization constants for 2-twist operators in twisted mass QCD}}, \href{https://doi.org/10.1103/PhysRevD.83.014503}{\emph{Phys. Rev. D} {\bfseries 83} (2011) 014503} [\href{https://arxiv.org/abs/1006.1920}{{\ttfamily 1006.1920}}].

\bibitem{Alexandrou:2015sea}
{\scshape ETM} collaboration, C.~Alexandrou, M.~Constantinou and H.~Panagopoulos, \emph{{Renormalization functions for Nf=2 and Nf=4 twisted mass fermions}}, \href{https://doi.org/10.1103/PhysRevD.95.034505}{\emph{Phys. Rev. D} {\bfseries 95} (2017) 034505} [\href{https://arxiv.org/abs/1509.00213}{{\ttfamily 1509.00213}}].

\bibitem{Alexandrou:2024ozj}
C.~Alexandrou, S.~Bacchio, J.~Finkenrath, C.~Iona, G.~Koutsou et~al., \emph{{Nucleon charges and $\sigma$-terms in lattice QCD}},  \href{https://arxiv.org/abs/2412.01535}{{\ttfamily 2412.01535}}.

\bibitem{Alexandrou:2019ali}
C.~Alexandrou et~al., \emph{{Moments of nucleon generalized parton distributions from lattice QCD simulations at physical pion mass}}, \href{https://doi.org/10.1103/PhysRevD.101.034519}{\emph{Phys. Rev. D} {\bfseries 101} (2020) 034519} [\href{https://arxiv.org/abs/1908.10706}{{\ttfamily 1908.10706}}].

\bibitem{Jay:2020jkz}
W.~I. Jay and E.~T. Neil, \emph{{Bayesian model averaging for analysis of lattice field theory results}}, \href{https://doi.org/10.1103/PhysRevD.103.114502}{\emph{Phys. Rev. D} {\bfseries 103} (2021) 114502} [\href{https://arxiv.org/abs/2008.01069}{{\ttfamily 2008.01069}}].

\bibitem{Neil:2022joj}
E.~T. Neil and J.~W. Sitison, \emph{{Improved information criteria for Bayesian model averaging in lattice field theory}}, \href{https://doi.org/10.1103/PhysRevD.109.014510}{\emph{Phys. Rev. D} {\bfseries 109} (2024) 014510} [\href{https://arxiv.org/abs/2208.14983}{{\ttfamily 2208.14983}}].

\bibitem{Alexandrou:2023qbg}
{\scshape Extended Twisted Mass} collaboration, C.~Alexandrou, S.~Bacchio, M.~Constantinou, J.~Finkenrath, R.~Frezzotti et~al., \emph{{Nucleon axial and pseudoscalar form factors using twisted-mass fermion ensembles at the physical point}}, \href{https://doi.org/10.1103/PhysRevD.109.034503}{\emph{Phys. Rev. D} {\bfseries 109} (2024) 034503} [\href{https://arxiv.org/abs/2309.05774}{{\ttfamily 2309.05774}}].

\bibitem{Galster:1971kv}
S.~Galster, H.~Klein, J.~Moritz, K.~H. Schmidt, D.~Wegener et~al., \emph{{Elastic electron-deuteron scattering and the electric neutron form factor at four-momentum transfers 5fm$^{-2} < q^2 < 14$fm$^{-2}$}}, \href{https://doi.org/10.1016/0550-3213(71)90068-X}{\emph{Nucl. Phys.} {\bfseries B32} (1971) 221}.

\bibitem{Ye:2017gyb}
Z.~Ye, J.~Arrington, R.~J. Hill and G.~Lee, \emph{{Proton and Neutron Electromagnetic Form Factors and Uncertainties}}, \href{https://doi.org/10.1016/j.physletb.2017.11.023}{\emph{Phys. Lett. B} {\bfseries 777} (2018) 8} [\href{https://arxiv.org/abs/1707.09063}{{\ttfamily 1707.09063}}].

\bibitem{ParticleDataGroup:2024cfk}
{\scshape Particle Data Group} collaboration, S.~Navas et~al., \emph{{Review of particle physics}}, \href{https://doi.org/10.1103/PhysRevD.110.030001}{\emph{Phys. Rev. D} {\bfseries 110} (2024) 030001}.

\bibitem{charges:2024pos}
C.~Alexandrou, S.~Bacchio, C.~Iona, G.~Koutsou, Y.~Li et~al., \emph{{Nucleon axial, tensor, and scalar charges and $\sigma$-terms in lattice QCD}}, {\emph{PoS} {\bfseries LATTICE2024} (2025) 316}.

\end{thebibliography}\endgroup

\end{document}